\documentclass[preprint]{aastex}
\usepackage{graphicx}
\usepackage{amsmath}
\usepackage[caption=false]{subfig}
\usepackage{natbib}
\usepackage{booktabs}
\usepackage{array}
\newcolumntype{L}{>{\centering\arraybackslash}m{7cm}}

\begin{document}

\title{Signature of a heliotail organized by the solar magnetic field and the role of non-ideal processes in modeled IBEX ENA maps: a comparison of the BU and Moscow MHD models}

\author{M. Kornbleuth\altaffilmark{1}}
\affil{\altaffilmark{1}Astronomy Department, Boston University, Boston, MA 
02215, USA} 
\email{kmarc@bu.edu}

\author{M. Opher\altaffilmark{1}}

\author{I. Baliukin \altaffilmark{2}$^{,}$\altaffilmark{3}$^{,}$\altaffilmark{4}}
\affil{\altaffilmark{2}Space Research Institute of Russian Academy of Sciences, Profsoyuznaya Str. 84/32, Moscow, 117997, Russia}
\affil{\altaffilmark{3}Moscow Center for Fundamental and Applied Mathematics, Lomonosov Moscow State University, GSP-1, Leninskie Gory, Moscow, 119991, Russia}
\affil{\altaffilmark{4}HSE University, Moscow, Russia}

\author{M. A. Dayeh \altaffilmark{5}$^{,}$\altaffilmark{6}}
\affil{\altaffilmark{6}Southwest Research Institute, P.O. Drawer 28510, San Antonio, TX 78228, USA}
\affil{\altaffilmark{7}University of Texas at San Antonio, San Antonio, TX 78249, USA}

\author{E. Zirnstein \altaffilmark{7}}
\affil{\altaffilmark{7}Department of Astrophysical Sciences, Princeton University, 
Princeton, NJ, 08544, USA}

\author{M. Gkioulidou \altaffilmark{8}}
\affil{\altaffilmark{8}Applied Physics Laboratory, Johns Hopkins University, 
Laurel, MD 20723, USA}

\author{K. Dialynas \altaffilmark{9}}
\affil{\altaffilmark{9}Office of Space Research and Technology, Academy of Athens, 10679 Athens, Greece}

\author{A. Galli \altaffilmark{10}}
\affil{\altaffilmark{10}Physics Institute, University of Bern, Bern, 3012, Switzerland}

\author{J. D. Richardson \altaffilmark{11}}
\affil{\altaffilmark{11}Kavli Institute for Astrophysics and Space Research and Department of Physics, Massachusetts Institute of Technology, Cambridge, MA, USA}

\author{V. Izmodenov\altaffilmark{2}$^{,}$\altaffilmark{3}$^{,}$\altaffilmark{12}}
\affil{\altaffilmark{12}Institute for Problems in Mechanics, Vernadskogo 101-1, Moscow, 119526, Russia}

\author{G. P. Zank \altaffilmark{13}}
\affil{\altaffilmark{13}Department of Space Science, The University of Alabama in Huntsville
4 Huntsville, AL 35805, USA}

\author{S. Fuselier\altaffilmark{14}}
\affil{\altaffilmark{14}Southwest Research Institute, San Antonio, TX, 78228, USA}

\begin{abstract}
Energetic neutral atom (ENA) models typically require post-processing routines to convert the distributions of plasma and H atoms into ENA maps. Here we investigate how two different kinetic-MHD models of the heliosphere (the BU and Moscow models) manifest in modeled ENA maps using the same prescription and how they compare with Interstellar Boundary Explorer (IBEX) observations. Both MHD models treat the solar wind as a single-ion plasma for protons, which include thermal solar wind ions, pick-up ions (PUIs), and electrons. Our ENA prescription partitions the plasma into three distinct ion populations (thermal solar wind, PUIs transmitted and ones energized at the termination shock) and models the populations with Maxwellian distributions. Both kinetic-MHD heliospheric models produce a heliotail with heliosheath plasma organized by the solar magnetic field into two distinct north and south columns that become lobes of high mass flux flowing down the heliotail, though in the BU model the ISM flows between the two lobes at distances in the heliotail larger than 300 AU. While our prescription produces similar ENA maps for the two different plasma and H atom solutions at the IBEX-Hi energy range (0.5 - 6 keV), the modeled ENA maps require a scaling factor of $\sim$2 to be in agreement with the data. This problem is present in other ENA models with the Maxwellian approximation of multiple ion species and indicates that a higher neutral density or some acceleration of PUIs in the heliosheath is required.
\end{abstract}


\section{Introduction}
One method for studying the global heliosphere is using energetic neutral atoms (ENAs) originating from the charge-exchange of an energetic ion with an ambient neutral atom (usually a proton and a neutral hydrogen (H) atom for the ENAs discussed in this paper). Launched in 2008, the Interstellar Boundary Explorer (IBEX) \citep{McComas09a} has observed ENAs from the heliosphere for over a full solar cycle \citep{McComas20}. IBEX has two ENA imagers, IBEX-Lo covering low ENA energies (0.01 - 2 keV; \citet{Fuselier09}) and IBEX-Hi covering higher energies (0.5-6 keV; \citet{Funsten09}), with central energies ranging from 0.71 to 4.29 keV. In this study, we only consider IBEX-Hi observations because of their high signal-to-noise ratio. Measured ENA intensities represent the line-of-sight integral of local proton intensities times the neutral H density, which means modeling is required to interpret the observations. To model ENAs in the heliosphere, global heliospheric solutions can be used to simulate ENA fluxes directly \citep{Heerikhuisen08} or can be used in post-processing to simulate ENA fluxes from the plasma and neutral distributions \citep{Prested08,Zirnstein13,Kornbleuth18,Baliukin20}. 

One method for modeling ENAs in the heliosphere requires partitioning the MHD plasma into multiple ion species of thermal solar wind ions and pick-up ions (PUIs). \citet{Zank10} showed that by partitioning the plasma into three populations that are processed at the termination shock (thermal solar wind, transmitted PUIs, and reflected PUIs) and modeling each as a Maxwellian distribution, one can reasonably replicate the total ion distribution that reflects IBEX-Hi energies. \citet{Zirnstein17} used the method of \citet{Zank10} and showed that extinction is also an important process in the heliosphere, where ions of a particular energy are depleted through charge exchange. This extinction process places a limit on how far observed ENAs can originate in the heliosphere, thereby restricting our line-of-sight in the heliotail.

\citet{Kornbleuth21} recently compared the MHD solutions from two different kinetic-MHD solutions, the Boston University (BU) model and the Moscow model, when both are simulated using identical boundary conditions and a kinetic treatment of neutrals. One key difference between the BU and Moscow models is that the BU model allows for non-ideal MHD effects, such as magnetic reconnection at the heliopause, while the Moscow model does not. This difference results in a more compressed heliosphere at high latitudes. Down the heliospheric tail, the interstellar plasma mixes with solar wind plasma along reconnected field lines approximately 300-400 AU from the Sun in the BU model, whereas the two fluids do not mix in the Moscow model. Besides these differences, the two solutions are quite similar. This paper explores whether these differences can be seen in ENA flux maps within the IBEX-Hi energy range.

\citet{McComas13} was the first to present IBEX ENA observations of the heliotail, and noted two high latitude lobes and two low latitude lobes that were all energy dependent. The high latitude north and south lobes appeared as regions of enhanced ENA flux at high energies ($>$2 keV), whereas the low latitude port and starboard lobes appeared as regions of relative low ENA flux at the same energies. \citet{Schwadron14} also presented IBEX ENA observations of the heliotail from the first five years of IBEX observations, but with the IBEX Ribbon removed via a masking and interpolation scheme to study the globally distributed flux (GDF), believed to primarily originate from the heliosheath. These GDF maps also showed a similar profile to those observed by \citet{McComas13}. \citet{Zirnstein17}, \citet{Kornbleuth20}, and \citet{Baliukin20} all produced modeled ENA maps of the heliosphere using different ENA models and different kinetic-MHD plasma solutions (with different boundary conditions) and were able to qualitatively reproduce the heliotail profile observed by IBEX-Hi. 

In this work, we study how two different heliospheric solutions affect the modeled ENA maps where the kinetic-MHD models use the same boundary conditions and the ENAs are modeled using the same recipe. In Section \ref{sec:Model}, we present summaries of both the kinetic-MHD models used and the ENA prescription. In Section \ref{sec:results}, we present the results of our investigation comparing the two kinetic-MHD solutions through modeled ENA maps. In Section \ref{sec:Summary}, we present a summary of our results and our conclusions.

\section{MHD and ENA Models} \label{sec:Model}
In this section, we discuss the two kinetic-MHD models (BU and Moscow) used in this work (Section \ref{ssec:mhd}) and the ENA model prescription (Section \ref{ssec:ena}). A more detailed description of the kinetic-MHD and ENA models can be found in \citet{Kornbleuth21} and \citet{Kornbleuth20}, respectively.

\subsection{MHD Models} \label{ssec:mhd}
For both MHD models, we use identical inner and outer boundary conditions from \citet{Izmodenov20}. In the interstellar medium, the proton density is assumed to be $n_{p,ISM}$ = 0.04 cm$^{-3}$, while the neutral H atom density is $n_{H,ISM}$ = 0.14 cm$^{-3}$. All the neutral and ionized populations in the interstellar medium are assumed to have the same bulk velocity $v_{ISM}$ = 26.4 km/s (longitude = 75.4$^{\circ}$, latitude = -5.2$^{\circ}$ in ecliptic J2000 coordinate system) and temperature $T_{ISM}$ = 6530 K at the outer boundary. We use the interstellar magnetic field intensity and orientation corresponding to $B_{ISM}$ = 3.75 $\mu$G and $\alpha$ = 60$^{\circ}$, where the magnetic field is aligned with the hydrogen deflection plane \citep{Lallement05} and $\alpha$ is the angle between the interstellar velocity and magnetic field vectors. 

The inner boundary conditions of the BU and Moscow models are implemented at 10 AU and 1 AU, respectively, and we match the two boundary conditions by extracting the solar wind conditions from the Moscow model at 10 AU and implementing them in the BU model. We use 22-year averaged solar cycle conditions (1995-2017) as in \citet{Izmodenov20}. Helio-latitudinal variations of the solar wind density and speed are taken into account, and the temperature is calculated from the speed by assuming a Mach number of M=6.44 at 1 AU (corresponding to a solar wind temperature of $T_{SW}$ = 188500 K at Earth's orbit) for all latitudes. Hourly-averaged solar wind data from the OMNI 2 dataset is used for the density and speed in the ecliptic plane, while heliolatitudinal variations of the solar wind speed are based on analysis of interplanetary scintillation (IPS) observations \citep{Tokumaru12} from 1990 to 2017. For heliolatitudinal variations of the solar wind mass flux, SOHO/SWAN full-sky maps of backscattered Lyman-alpha intensities are used \citep{Quemerais06, Lallement10,Katushkina13, Katushkina19}. For the solar magnetic field, a Parker solution is assumed for both models, with the radial component of the magnetic field as $B_{r,SW}$ = 37.5 $\mu$G at 1 AU. A unipolar configuration of the solar magnetic field is used because the unipolar treatment eliminates spurious numerical effects due to numerical diffusion and reconnection of the solar magnetic field across the heliospheric current sheet \citep{Michael18}.

The BU model uses the Solar-wind with Hydrogen Ion Exchange and Large-scale Dynamics (SHIELD) numerical model based on the works of \citet{Michael19,Michael21}. This model extends the multi-fluid MHD model used in \citet{Opher15} (first implemented in the BU model in \citet{Opher09}) to include a kinetic treatment of neutral H atoms instead of a multi-fluid treatment of H atoms \citep{Zank96}. The BU model uses the Space Weather Modeling Framework (SWMF) \citep{Toth05} and couples the Outer Heliosphere (OH) and Particle Tracker (PT) components. The PT component of SWMF is based on the Adaptive Mesh Particle Simulation (AMPS) \citep{Tenishev21} to treat neutral atoms kinetically by solving the Boltzmann equation for neutral H atoms streaming through the domain and only incorporating effects due to charge exchange. Photoionization is not included, which is a negligible effect at 10 AU where the inner boundary is located. The OH component is based on the Block-Adaptive Tree Solar wind Roe-Type Upwind Scheme (BATS-R-US) solver \citep{Toth12}, which is a three-dimensional (3D), block adaptive, upwind finite-volume MHD code. The BU model couples the PT component to the OH component, and iterates between the two components until a steady state solution is achieved.

The Moscow model treats the partially ionized interstellar plasma as a two-component gas consisting of neutral H atoms and a charged plasma consisting of protons and electrons and helium ions (He+). In the solar wind, the plasma consists of protons, electrons, and alpha particles (He++). Here, He+ and He++ are neglected for the purposes of comparison with the BU model. The neutral atoms are treated kinetically, while the plasma is described via the ideal MHD equations (the velocity distribution of the proton component is assumed to be locally Maxwellian). Source terms for the MHD equations are calculated via the kinetic treatment of neutrals, and integrals of the H-atom velocity distribution are calculated using a Monte Carlo method \citep{Malama91} which solves the Boltzmann transport equation (e.g. \citet{Izmodenov01}). A global iteration method \citep{Baranov93} gives a self-consistent steady-state solution. A 3D moving grid is used to fit discontinuities. A fitting technique proposed by \citet{Godunov79} allows them to fit all the major discontinuities present in the simulation – the heliopause, termination shock and bow shock (when applicable; in this run there is no bow shock).

In the BU model the solar and interstellar magnetic field lines are able to reconnect, but in the Moscow model no communication is allowed at the heliopause. The heliopause in the BU model allows for outflow of the solar wind plasma due to reconnection in the tail. The heliopause in the Moscow model does not allow for this reconnection, and forms a long tail in accordance with studies by \citet{Korolkov21}. The unipolar magnetic field allows reconnection to occur in the port side and tail of the BU model while minimizing magnetic dissipation within the heliosheath. \citet{Opher17,Opher20} suggest that this reconnection explains the draped magnetic field ahead of the heliopause as revealed by Voyager 1 and 2 data. Using a dipolar solar magnetic field, \citet{Michael18} showed that reconnection across the current sheet with the interstellar magnetic field at the heliopause leads to a draped interstellar magnetic field that instead deviates from observations. \citet{Pogorelov15} suggested that in the BU model the inclusion of a bipolar solar magnetic field configuration, kinetic neutrals, or solar cycle would suppress the split-tail feature. \citet{Michael18} modeled the solar magnetic field both with a unipolar and bipolar configuration in a multi-fluid neutral treatment and found the split-tail persisted for the BU model. In a later work, \citet{Michael21} modeled the solar magnetic field with a unipolar configuration treating the neutrals kinetically and concluded a split tail was present in the BU model as well. Despite the inclusion of a bipolar solar magnetic field and kinetic neutrals treatment, the BU model has not been modeled with solar cycle conditions, so further investigation on whether the split tail remains is required.

\subsection{ENA Model} \label{ssec:ena}
The ENA Flux model is the same as used in \citet{Kornbleuth20}, which is an update to the model used in \citet{Kornbleuth18}. We
interpolate our Cartesian grid to a spherical grid of 2 AU $\times$ 6$^{\circ}$ $\times$ 6$^{\circ}$. From our model, we extract bulk plasma velocity streamlines to model ions crossing the termination shock and transiting through the heliosphere before charge exchanging to become ENAs. As in \citet{Zank10}
and \citet{Zirnstein17}, we partition the total thermal energy of the plasma via

\begin{equation}
T_{p}=\left(\frac{n_{\mathrm{SW}}}{n_{p}}\Gamma_{\mathrm{SW}}+
\frac{n_{\mathrm{tr}}}{n_{p}}\Gamma_{\mathrm{tr}}+\frac{n_{\mathrm{ref}}}
{n_{p}}\Gamma_{\mathrm{ref}}\right)T_{p},
\label{eq:Zank}
\end{equation}

\noindent where $n_{p}$ and $T_{p}$ are the density and temperature of the 
plasma, $n_{ip}$ is the density for the respective ion population, and $\Gamma_{ip}$ is 
the temperature fraction for the respective ion population given by $\Gamma_{ip}=T_{ip}/T_{p}$, with 
$T_{ip}$ being the temperature for the respective ion population. We divide the plasma into three distinct ion populations:
thermal solar wind ions, PUIs created in the supersonic solar wind which are adiabatically transmitted across the termination shock 
(transmitted PUIs) and PUIs created in the supersonic solar wind which are reflected at the termination shock until they have sufficient energy to overcome the cross-shock potential (reflected PUIs) \citep{Zank96, Zank10}. As in \citet{Kornbleuth20}, we use density ratios relative to the plasma given by 0.836, 0.151, and 0.013 for the thermal ions, transmitted PUIs, and reflected PUIs, respectively, and energy ratios ($n_{ip}\Gamma_{ip}$) given by 0.04, 0.50, and 0.46 for the direction of the nose. We keep the energy ratio constant for all locations along the termination shock, but we vary the density ratios along the termination
shock in different directions by using the works of \citet{Lee09} and \citet{Zirnstein17}, where the PUI fraction can be determined using

\begin{equation}
\alpha(\mathbf{r}_{TS})=\frac{r_{TS}}{u_{1 AU}n_{1 AU}}<n_{H}>
(\nu_{ph}(1 AU)+\sigma_{ex}u_{1 AU}n_{1 AU}).
\label{eq:puirat}
\end{equation}

\noindent Here, $r_{TS}$ is the distance to the termination shock, $u_{1 AU}$ is the speed of the solar wind at 1 AU, $n_{1 AU}$ 
is the density of the solar wind at 1 AU, $\sigma_{ex}$ is the charge exchange cross-section from \citet{Lindsay05}, and $<n_{H}>$ is the average neutral H density between the inner boundary and the termination shock for a given direction. The term $\nu_{ph}(1 AU)$ is the photoionization rate at 1 AU assumed to be a constant value of $8\times10^{-8}$ s$^{-1}$ in all directions, even though it realistically varies in time and space (e.g., \citealt{Sokol19}). We can use the
PUI fraction, $\alpha$, to determine how the density and energy fractions for the thermal ions, transmitted PUIs, and reflected PUIs vary
in all directions at the termination shock via

\begin{equation}
n_{i}(\mathbf{r})=n_{ip}(\mathbf{r})\frac{\alpha(\theta,\phi)}
{\alpha(\theta_{nose},\phi_{nose})}e^{-\xi(\mathbf{r})},
\label{eq:ang_pui}
\end{equation}

\noindent where $n_{i}$ is the ion density for a given direction when the variation in $\alpha$ at termination shock is considered, $\theta$ is the polar angle, and $\phi$ is the azimuthal angle. The polar angle increases from the northern pole toward the southern pole, while the azimuthal angle increases in the clockwise direction from the nose. The parameter $n_{ip}$ is the density of the PUI species for a given direction calculated using the total plasma density multiplied by the density ratios at the nose mentioned above.

As ions of a particular energy travel along a given streamline, some will undergo charge exchange and the newly created PUIs will have a different
energy. This ``extinction'' by charge exchange limits the distance to which we can observe ENAs created past the termination shock, with the
distance out to which 1/$e$ of ions of a particular energy remain referred to as the cooling length. This process can be calculated
along streamlines using the method of \citet{Zirnstein17} such that the extinction along a streamline is given by

\begin{equation}
\xi(\mathbf{r})=\int_{r_{TS}}^{r} \frac{n_{H}(\textbf{r})\sigma_{ex}(E)v_{i}(E)ds}{u_{p}(\textbf{r})}.
\label{eq:ext}
\end{equation}

\noindent Here, $r_{TS}$ is the streamline distance to the TS, $n_{H}$ is the neutral H density, $v_{i}$ is the speed of the parent proton which will yield an ENA of a particular energy, $u_{p}$ is the bulk plasma speed, and $ds$ is the path length over which we integrate the streamline. We define the energy ($E$) by the parent ion energy because at IBEX-Hi energies the parent ion velocity is much larger than the velocity of an interstellar neutral ($\mathbf{v_{i}} \gg \mathbf{v_{H}}$). The newly formed heliosheath ions which replace the extinguished ions have a low characteristic energy ($\sim$0.1 keV), and because we are focusing on higher energy ENAs to which these newly formed heliosheath ions will have little contribution, we do not include them in our ENA modeling. 

We calculate the ENA flux along a radial line-of-sight (LOS) using

\begin{equation}
J(E,\theta,\phi)=\int_{r_{observer}}^{\infty}\frac{2E}{m_{p}^{2}}f_{p}(n_{i}(\textbf{r}'(s)),T_{i}(\textbf{r}'(s)),v_{plasma}(\textbf{r}'(s)))n_{H}(\textbf
{r}'(s))\sigma_{ex}(E)S(E)d\textbf{r}'(s),
\label{eq:flux}
\end{equation}

\noindent where $\textbf{r}'$ is the vector along a particular LOS as a function of $\theta$ and $\phi$, $s$ is the distance along the vector, $m_{p}$ is the mass of a proton and $f_{p}$ is the phase space velocity distribution, which is treated as a Maxwellian for each modeled ion population. The velocity of the parent ion in the frame of the plasma is given by $v_{plasma}=|\mathbf{v_{p}}-\mathbf{v_{i}}|$, with $\mathbf{v_{p}}$ and $\mathbf{v_{i}}$ being the velocities of the bulk plasma and the parent proton, respectively. For the density and temperature of the given ion population, we use $n_{i}$ and $T_{i}$, respectively. $T_{i}$ is a fraction of the local MHD temperature based on the thermal pressure fraction of the ion species. We assume quasi-neutrality and use the approximation that the electrons have the same temperature as the solar wind ions. Therefore, the plasma temperature is given by \citep{Zirnstein17}

\begin{equation}
T_{p}=\frac{2T_{\mathrm{MHD}}}{1+\Gamma_{SW}},
\end{equation}

\noindent where $T_{\mathrm{MHD}}$ is the temperature given from the MHD solution. We can calculate the temperature of a particular ion 
population using the energy and density ratios of the population with respect to the plasma.

We also include the survival probability of ENAs in our model, shown by \citet{Bzowski08}. The survival probability represents the likelihood that an ENA
created in the heliosheath will charge exchange prior to being observed by IBEX. Unlike in the calculation for ion extinction, we calculate
the survival probability of an ENA along a radial LOS, since the trajectories of H atoms with IBEX-Hi energies are almost straight. The survival probability is given by 

\begin{equation}
S(E)=\int_{r_{source}}^{r_{observer}} 
\frac{\sigma(v_{rel})v_{rel}n_{p}}{v_{ENA}}dr,
\label{eq:surv}
\end{equation}

\noindent where dr is the radial element over which we are integrating, $v_{ENA}$ is the speed of the ENA, and $v_{rel}$ is the relative velocity
between the ENA and the bulk plasma given by \citep{Heerikhuisen06},

\begin{equation}
v_{rel}=v_{th,p}\left[\frac{e^{-\omega^{2}}}{\sqrt{\pi}}+\left(\omega+\frac{1}{2\omega}\right)erf(\omega)\right], \omega=\frac{1}{v_{th,p}}{\vert}\mathbf{v_{ENA}}-\mathbf{u_{p}}{\vert}.
\label{eq:rel}
\end{equation}

\noindent Here, $\mathbf{v_{ENA}}$ is the velocity of the ENA, $\mathbf{u_{p}}$ is the bulk averaged plasma velocity, and $v_{th,p}$ is the thermal 
speed of the plasma. Unlike in Equations \ref{eq:ext} and \ref{eq:flux}, we need to use $v_{rel}$ since the velocities of the ENA and bulk plasma are within an order of magnitude. The background plasma distribution is assumed to have a Maxwellian distribution, a simplified assumption for the multi-Maxwellian proton distribution in the heliosheath. We place the observer at the termination shock, similar to IBEX ENA maps, which utilizes a survival probability correction out to 100 AU. Because we only calculate the survival probability to the termination shock, where photoionization is negligible, we only include ionization via charge exchange in our calculation.

\section{Results}\label{sec:results}
\begin{figure*}[t!]
\centering
  \includegraphics[scale=0.45]{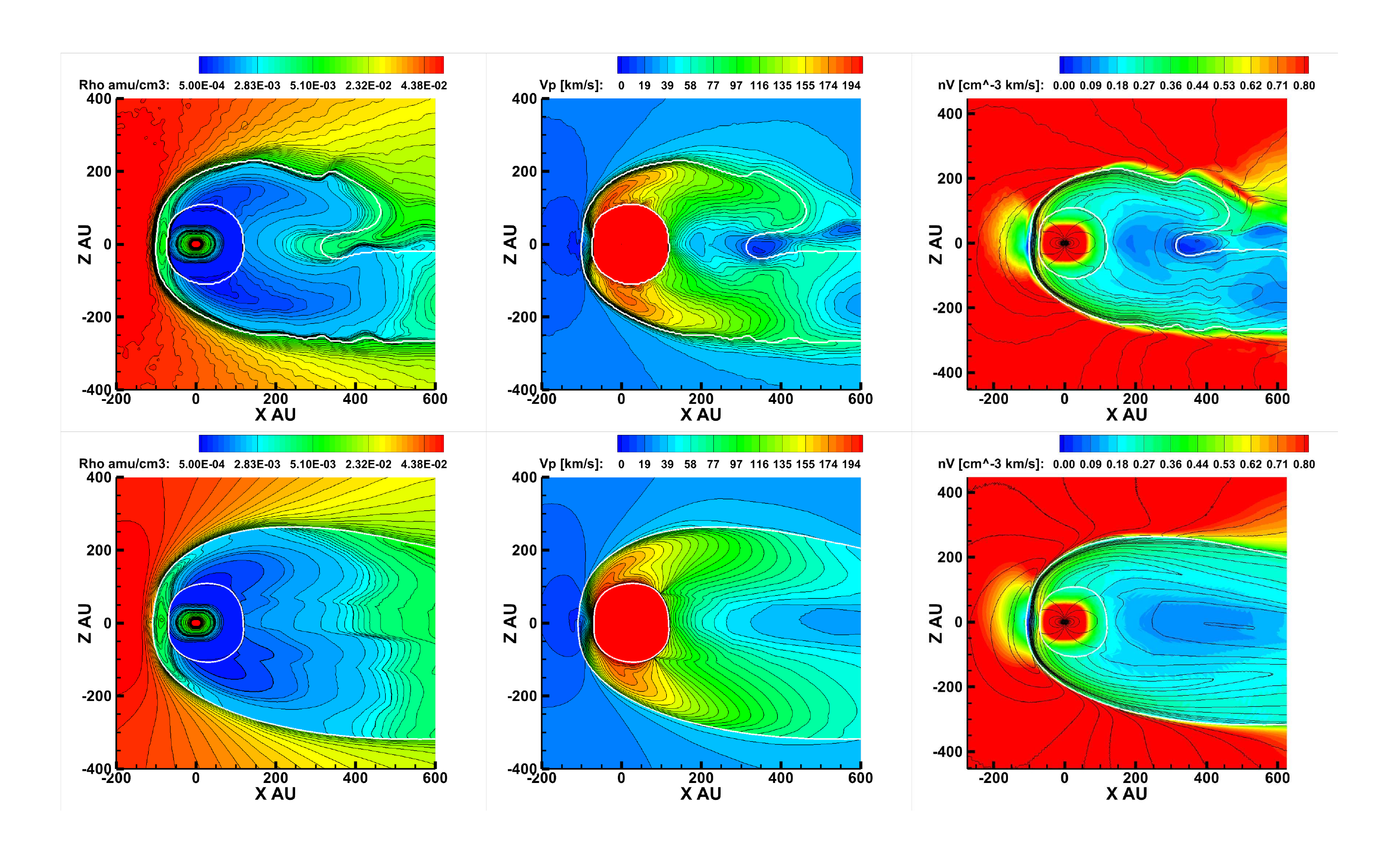}
  \caption{Meridional slices of plasma conditions within the heliosheath for the BU model (top) and the Moscow model (bottom). Presented are color and line contours of the plasma density [cm$^{-3}$] (left column) and the plasma speed [km/s] (middle). We also include the mass flux (right column) of each model, with the line contours corresponding to the magnetic field intensity. The white lines identify the termination shock (inner) and heliopause (outer) for each model.}
  \label{fig:heliosheath}
\end{figure*}

Both models produce a heliotail where the heliosheath flow (and therefore the PUIs) is organized by the solar magnetic field and concentrated into northern and southern lobes of higher mass flux compared to low latitudes \citep{Kornbleuth21}. The main difference between the models (Figure \ref{fig:heliosheath}) is that the ISM flows between the two lobes at distances larger than 300 AU down the tail in the equatorial plane in the BU model. At these ENA energies, \citet{Schwadron14} used average flow parameters to estimate the cooling length (the distance over which there are sufficient PUIs to generate ENAs of a particular energy before being depleted by charge exchange) to be on the order of 100-130 AU beyond the termination shock depending on the energy range. Within these distances of the termination shock, the two heliospheric solutions of the BU and Moscow MHD models are qualitatively similar (Figure \ref{fig:heliosheath}).

There are other minor differences between the MHD solutions due to the treatment of the heliopause boundary. As shown in \citet{Kornbleuth21}, there is magnetic reconnection between the solar and interstellar magnetic field at the heliopause, which is not present in the Moscow model. The magnetic reconnection in the BU model leads to an increased magnetic pressure outside the heliopause due to the twisting of the interstellar magnetic field, which compresses the BU modeled heliosphere relative to the Moscow model. This compression is noted in the inward motion of the heliopause, most notably at high latitudes.

In Figure \ref{fig:maps}, we present a comparison of modeled global ENA fluxes from the BU and Moscow models with IBEX GDF observations averaged over the years 2009 to 2013 \citep{Schwadron14}. The BU and Moscow models are multiplied by a factor of 1.8 for comparison with the IBEX data. In comparing the ENA maps at all energies, the maps appear qualitatively similar. 

At 1.11 keV, both the MHD models and the IBEX data have enhanced ENA flux at lower latitudes in the nose and tail directions, while there is less flux at high latitudes. At the higher energies (2.73 and 4.29 keV), there are two distinct lobes which appear at high latitudes in the tail, whereas there is a deficit of flux at low latitudes in the flanks, referred to as port and starboard lobes. The similarity of the maps captures both the effect of the fast solar wind in the poles and the slow solar wind in the equatorial plane \citep{McComas13}, as well as the heliotail where the solar magnetic field organizes the solar particles (and PUIs) into two distinct north and south columns \citep{Opher15,Drake15,Kornbleuth20}. 

Despite the similarities between the models and the observations, there are some differences in the global ENA maps. At the 1.11 keV energy band, we find the region of enhanced ENA flux in the tail to reach higher latitudes (by 15-30$^{\circ}$ in the north and south) in both models than is seen in the IBEX GDF. Additionally, there is a gap near the downwind direction where both model fluxes decrease, which is a feature not present in the IBEX GDF where the ENA flux peaks around the downwind direction. In the 2.73 keV energy band, the north and south lobes that are produced in the BU and Moscow models have higher ENA fluxes relative to all other regions of the sky than are seen in the IBEX GDF data, and the BU and Moscow model ENA intensities diverge at 4.29 keV. We note again that both models are scaled up by a factor of 1.8, and therefore they still underestimate the observed fluxes by a finite amount.

As in \citet{Kornbleuth20}, we need to scale modeled ENA maps by a factor of 1.8 to provide an average global agreement with observations. In other works, such as \citet{Zirnstein17} and \citet{Shrestha21} which also use the superposition of Maxwellian distributions for the ions from \citet{Zank10}, a scaling factor of 2.5 and $\sim$3 is required as well, respectively. \citet{Baliukin20} use a kinetic approach to PUI modeling in post-processing and assume that PUIs form a filled shell distribution downstream of the TS. They do not require scaling of their results in comparison with IBEX observations at $\sim$1-2 keV, but underestimate the ENA flux at the lowest and highest energy channels of IBEX-Hi due to the filled shell distribution assumption. The quantitative discrepancy with observations may result from the following facts: (1) the interstellar neutral H density is higher than it was assumed, as recently reported by \citet{Swaczyna20} to be 0.127 $\pm$ 0.015 cm$^{-3}$ upwind at the termination shock; (2) the ENA prescription used in this paper neglects the adiabatic heating of PUIs due to compression of plasma in the inner heliosheath; (3) some acceleration of PUIs in the heliosheath is ignored. With respect to the higher interstellar H density found by \citet{Swaczyna20}, their estimated value is greater than the value upwind at the termination shock in the BU and Moscow models by a factor of $\sim$1.25, which can partially account for the discrepancy.

The modeled ENA maps using the solutions of the two kinetic-MHD models are very similar although there are differences. At all energies, the ENA flux in the nose region of the heliosphere of the two MHD models matches well. Both have regions of high ENA flux at lower latitudes in the 1.11 keV energy band. At higher energies at high latitudes, both models produce north and south lobes in the heliotail of high ENA flux, as compared to lower latitudes where there is a deficit of flux in the form of port and starboard lobes as is seen in IBEX observations. 

\begin{figure*}[t!]
\centering
  \includegraphics[scale=0.5]{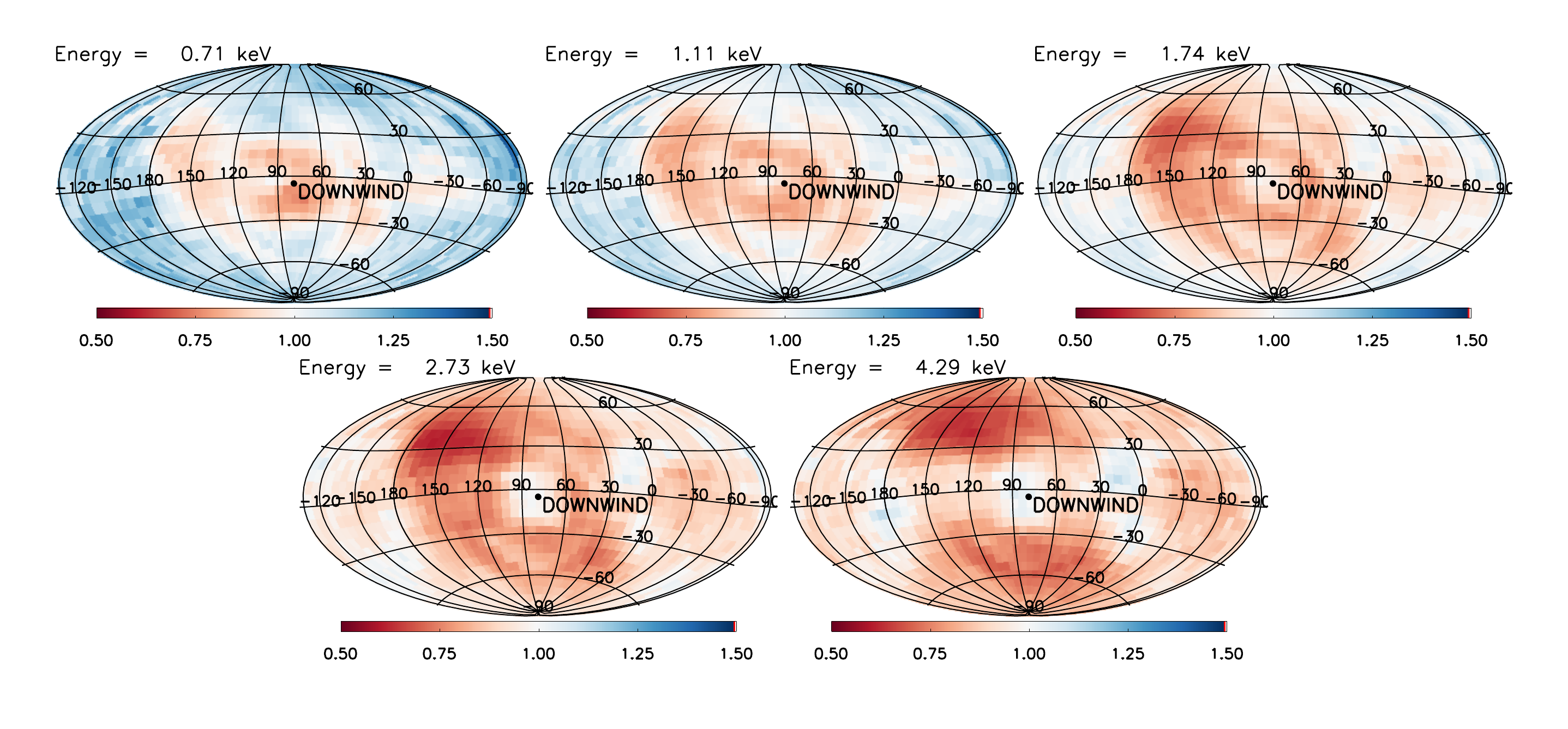}
  \caption{Ratios of ENA Flux from the Moscow model relative to the BU model for the 0.71 keV (top left), 1.11 keV (top middle), 1.74 keV (top right), 2.73 keV (bottom left), and 4.29 keV (bottom right) energy bands.}
  \label{fig:mapratios}
\end{figure*}

While there is general agreement between the BU and Moscow model results, there are differences between the models in the downwind hemisphere resulting primarily from the higher plasma density in the heliosheath of the BU model. Figure \ref{fig:mapratios} shows the ratio of the ENA flux from the Moscow model relative to the BU model and we find that the ENA flux is higher in the BU model than in the Moscow model in the downwind hemisphere. In particular at the 2.73 and 4.29 keV energy bands, the ENA flux present in the high latitude lobes in the tail region of both models is higher in the BU model than in the Moscow model. For all energies, there is good agreement between the modeled ENA flux in the upwind hemisphere of the heliosphere where the solutions are similar.

\begin{figure*}[t!]
\centering
  \includegraphics[scale=0.6]{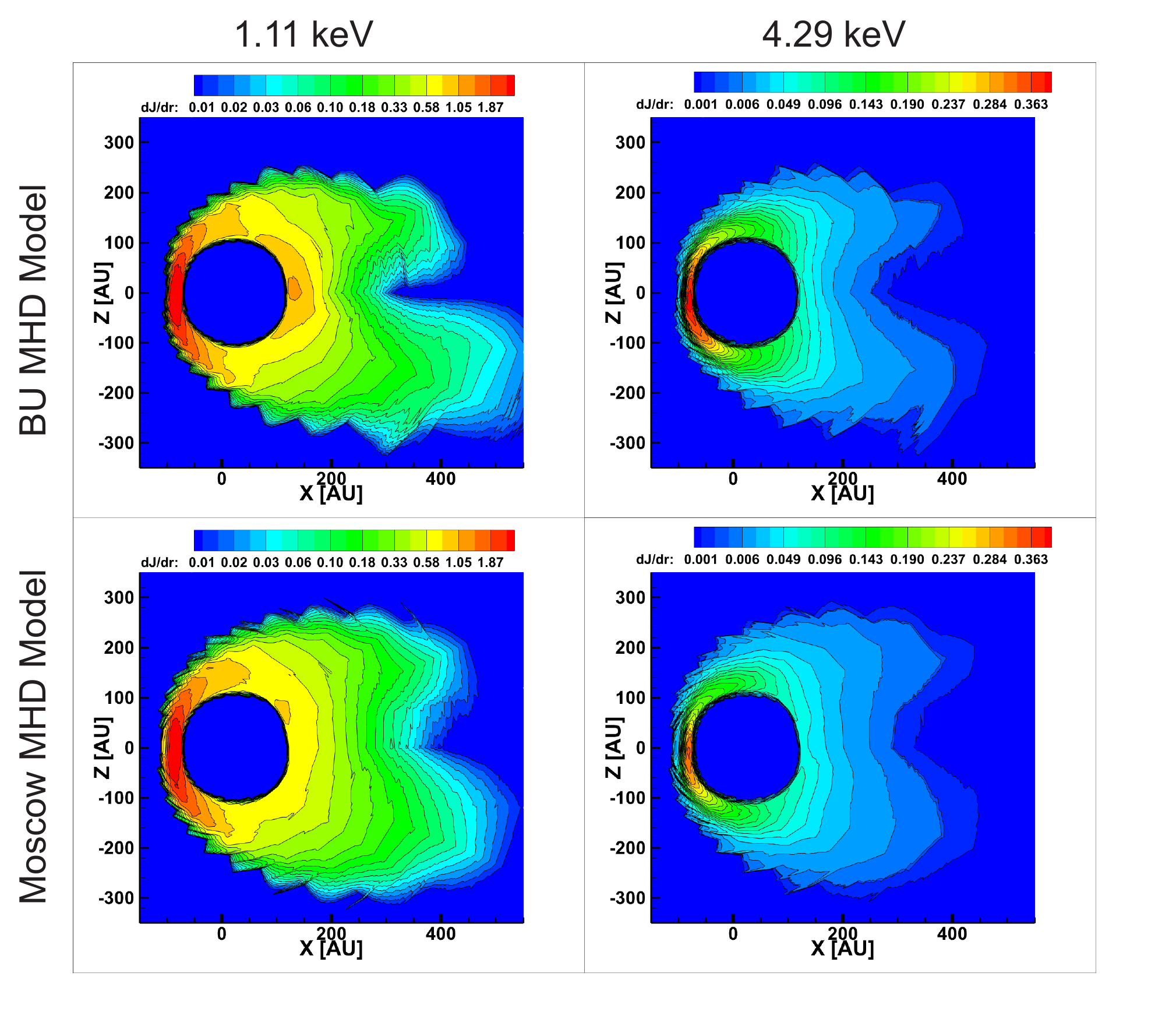}
  \caption{Meridional slices of ENA flux production (dJ/dr)with units of (cm$^{3}$ s sr keV)$^{-1}$ for the BU model (top) and Moscow model (bottom). Included is the flux production at energies of 1.11 keV (left) and 4.29 keV (right).}
  \label{fig:meridional}
\end{figure*}

\begin{figure*}[t!]
\centering
  \includegraphics[scale=0.4]{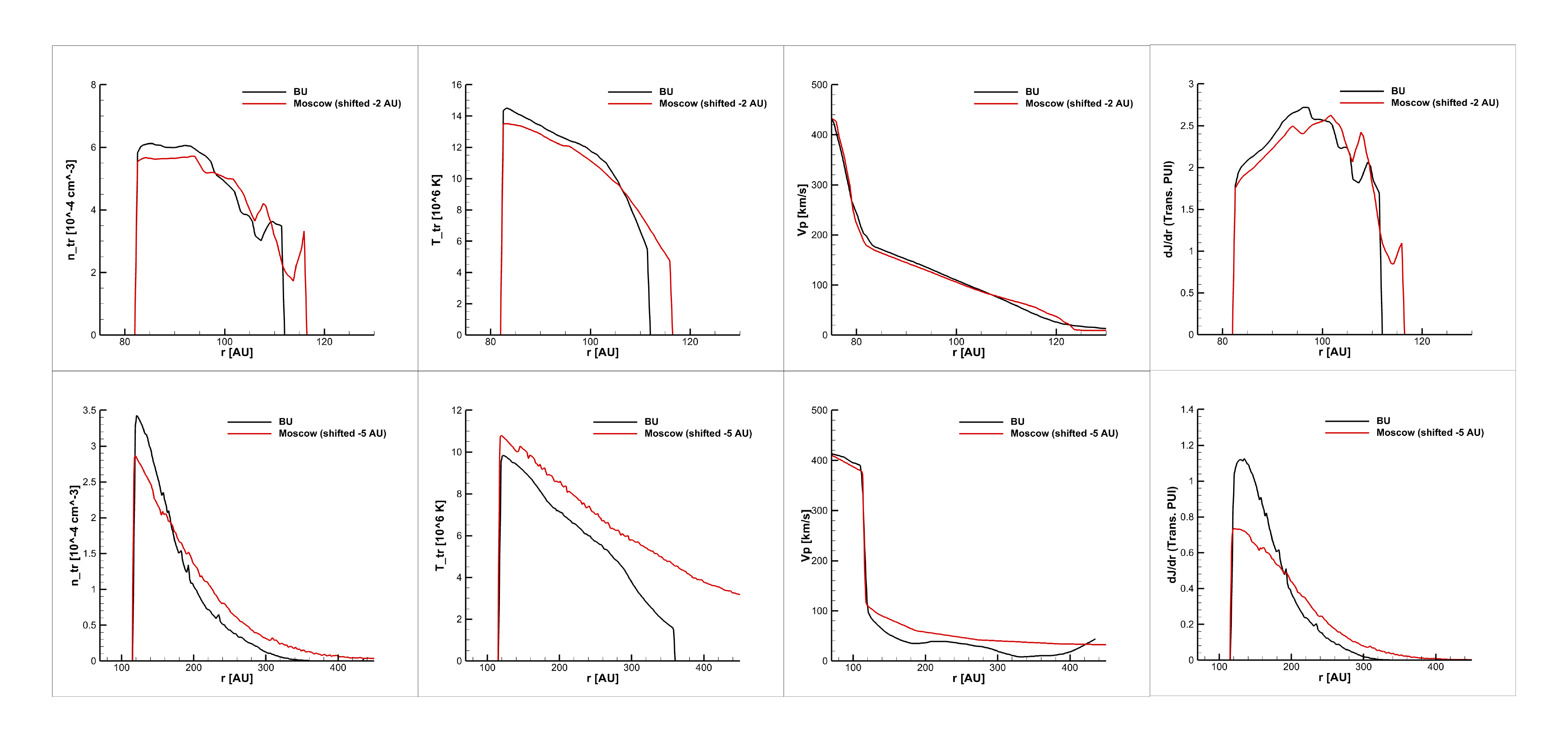}
  \caption{1D cuts along the Voyager 1 (top, A-D) and downwind (bottom, E-H) directions showing the contribution of different ions properties to ENA flux production for the 1.11 keV energy band. Included is the transmitted PUI density (cm$^{-3}$) (A \& E), the transmitted PUI temperature (K) (B \& F), the plasma speed (km/s) (C \& G), and ENA flux production from transmitted PUIs (cm$^{3}$ s sr keV)$^{-1}$ (D \& H). Included are results from the Moscow model (red) and the BU model (black). The modeled ENA flux at 1.11 keV in the Voyager 1 direction is 35.5 and 39.4 (cm$^{2}$ s sr keV)$^{-1}$ from the BU and Moscow models, respectively. In the downwind direction, the modeled flux at 1.11 keV is 49.3 and 41.4 (cm$^{2}$ s sr keV)$^{-1}$ for the BU and Moscow models, respectively.}
  \label{fig:1d_1kev}
\end{figure*}

\begin{figure*}[t!]
\centering
  \includegraphics[scale=0.4]{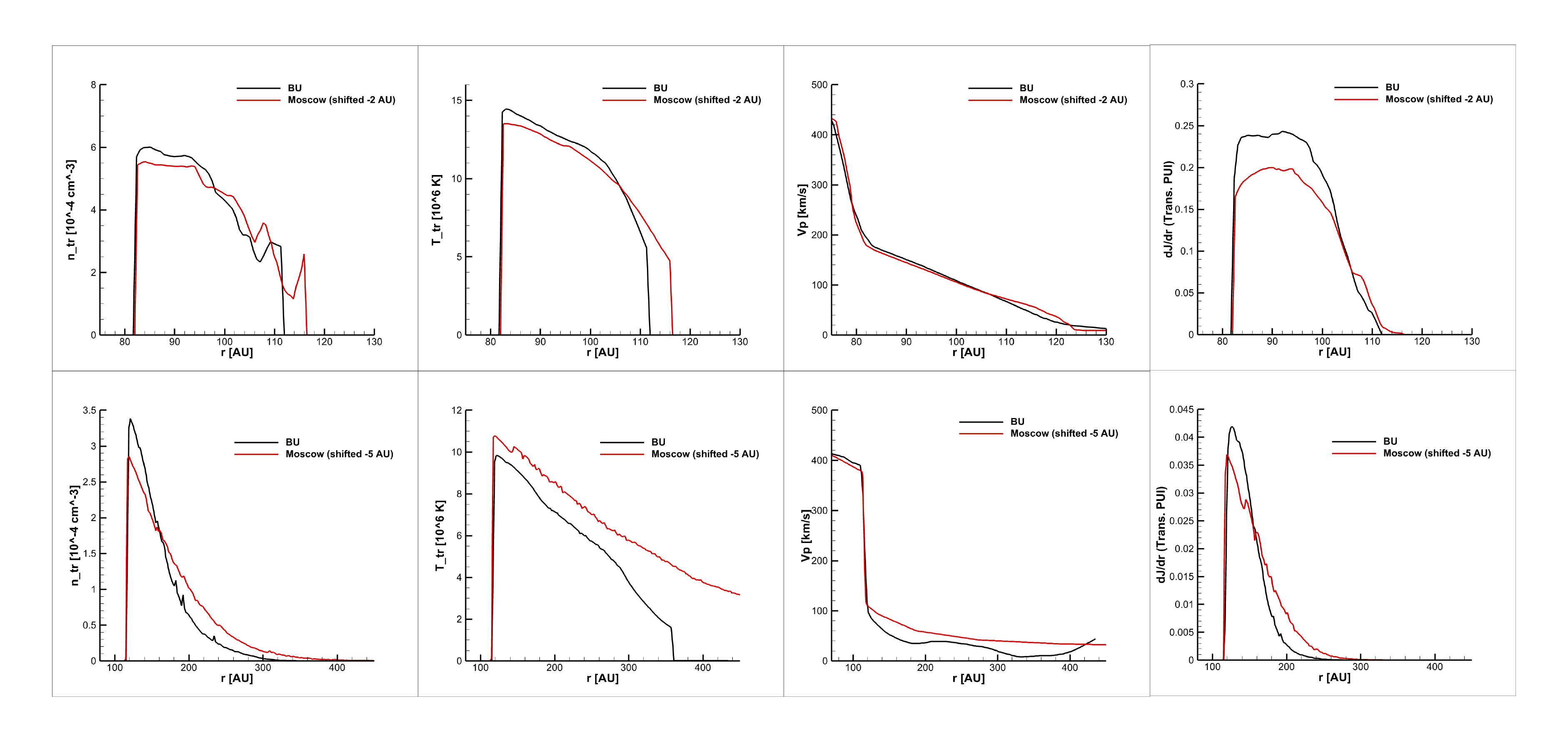}
  \caption{Same as Figure \ref{fig:1d_1kev}, but for the 4.29 keV energy band. The modeled ENA flux at 4.29 keV in the Voyager 1 direction is 3.7 and 3.4 (cm$^{2}$ s sr keV)$^{-1}$ from the BU and Moscow models, respectively. In the downwind direction, the modeled flux at 4.29 keV is 2.2 and 2.1 (cm$^{2}$ s sr keV)$^{-1}$ for the BU and Moscow models, respectively.}
  \label{fig:1d_4kev}
\end{figure*}

A comparison of the ENA flux production in the meridional plane is presented in Figure \ref{fig:meridional} for the 1.11 keV and 4.29 keV energy bands. We find that for both models that ENA flux production is similar at the 1.11 and 4.29 keV energy bands. Due to the higher ENA production with less extinction at lower energies (1.11 keV), there is still a small contribution at distances $>$300 AU where the different heliotail configurations begin to manifest, yet the ENAs created beyond this distance contribute only a small amount to the overall ENA flux. In contrast, at the higher energies (4.29 keV), there is more extinction than at lower energies and there is effectively no contribution at distances where the heliotail differences are present in the models. This can be seen in Figures \ref{fig:1d_1kev} and \ref{fig:1d_4kev}.

Figures \ref{fig:1d_1kev} and \ref{fig:1d_4kev} show line cuts along the Voyager 1 and downwind directions for the BU and Moscow models at the 1.11 and 4.29 keV energies. We present the transmitted PUI density, temperature, and speed in the heliosheath, as well as the ENA flux production from the transmitted PUIs at these energies (note that transmitted PUIs contribute the most to the IBEX-Hi ENA spectrum). The Voyager 1 location is chosen as it represents a location in the nose region of the heliosphere where the IBEX Ribbon does not contribute a significant amount to the total observed ENA flux. We also choose the downwind location to demonstrate how the low latitude tail of both models produces similar ENA profiles, even if the BU model has a higher ENA flux.

At 1.11 keV (Figure \ref{fig:1d_1kev}), both the Voyager 1 and downwind ENA flux is mediated by the transmitted PUIs. For both models, the flow speed of the ions is similar in the Voyager 1 direction, but the BU model has a lower flow speed in the downwind direction. Flow speed is an important quantity for lower energy ($<$2 keV) ENAs, so the lower flow speed at the termination shock in the BU model leads to an increase in ENA fluxes observable at 1 AU. The BU model has a slightly denser and hotter PUI population downstream of the termination shock. While the transmitted PUI density plays the primary role for differences in the models' flux production, the cooler PUI temperature in the Moscow model leads to more ENA production at lower energies compared to higher energies. Therefore, in the Voyager 1 direction the total integrated flux is similar between the two models since the slightly cooler PUI temperature of the Moscow model offsets the slightly denser PUIs in the BU model. In contrast, the longer heliotail in the Moscow model does not lead to a significantly different ENA flux in the downwind direction, since the cooling length restricts the ENA production to within $\sim$100 AU of the termination shock. 

At 4.29 keV (Figure \ref{fig:1d_4kev}), the difference in transmitted PUI densities between the two models also plays a significant role in their ENA flux production, yet unlike in the 1.11 keV energy band, the temperature also plays a role in mediating the flux production, as can be seen in the case of Voyager 1. For this particular direction, the BU model has a higher transmitted PUI density and temperature in the heliosheath, while the Moscow model has a slightly thicker heliosheath. The hotter and denser PUIs in the Voyager 1 direction of the BU model lead to a higher ENA flux in the BU model at energies $>$2 keV. This effect is less apparent in the downwind direction, where the PUIs in the Moscow model have a hotter temperature, yet the ENA flux production more closely tracks the transmitted PUI density. One reason why the transmitted PUI temperature has less of an effect in the downwind direction as compared to the Voyager 1 direction is because the plasma, and therefore the transmitted PUIs, is cooler in the downwind direction than in the Voyager 1 direction. 

In the downwind direction, the cooling length restricts the line-of-sight such that the differences in the length of the heliotail cannot be distinguished. We find the cooling length ($l_{c}$) to be $l_{c} = 64$ AU (178 AU from the Sun) and $l_{c} = 97$ AU (216 AU from the Sun) for the BU and Moscow models at 1.11 keV, respectively. At the 4.29 keV energy, we find the cooling length to be $l_{c} = 54$ AU (168 AU from the Sun) and $l_{c} = 79$ AU (198 AU from the Sun) for the BU and Moscow models, respectively. For both models and both energies, the cooling length is located within 300 AU of the Sun, where the the tail solutions begin to deviate as ISM plasma mixes with solar wind plasma in the BU model. While the plasma solutions are similar within the cooling length distances, as noted in Figures \ref{fig:1d_1kev} and \ref{fig:1d_4kev}, the plasma speed in the downwind direction is notably lower in the BU model than the Moscow model, which contributes to a more inward cooling length due to the slower flow allowing for more charge exchange events over shorter distances.

\begin{figure*}[t!]
\centering
  \includegraphics[scale=0.55]{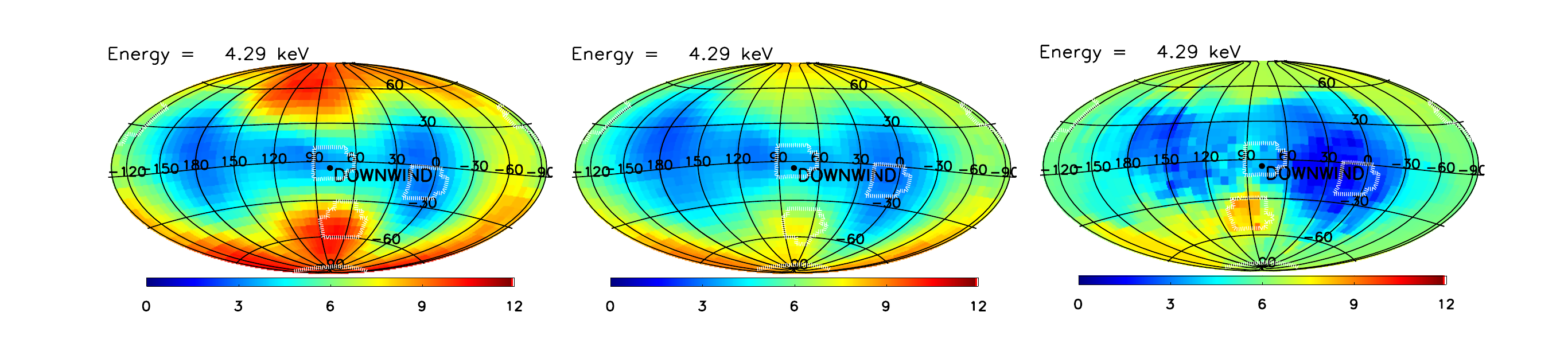}
  \caption{Regions of sky used for comparison in Figure \ref{fig:fluxcomp}. Regions are outlined in white on top of ENA sky map of flux centered on the downwind (tail) direction in units of (cm$^{2}$ s sr keV)$^{-1}$ for the BU model (left), Moscow model (middle), and IBEX GDF from 2009-2013 (right) in the 4.29 keV energy band. From the north pole downward, the regions reflect the directions of Voyager 1, downwind, the port lobe, the south lobe, and the south pole. The BU and Moscow fluxes are multiplied by a factor of 1.8.}
  \label{fig:regions}
\end{figure*}

To further investigate, we compare the ENA flux in five different directions of interest. Unlike in Figure \ref{fig:maps} where we present a comparison with IBEX GDF data averaged over the first five years of IBEX observations with the ribbon removed to approximate heliosheath flux from \citet{Schwadron14}, here we compare with IBEX observations where the ribbon is still present. We use IBEX-Hi, annually combined, ENA sky maps during 5 years between 2009 January and 2013 December at five energy passbands centered at 0.7, 1.1, 1.7, 2.7, and 4.3 keV. Data has been acquired from the validated Data Release \#16 \citep{McComas20} and is accessible at https://ibex.princeton.edu/DataRelease16. Data is survival probability-corrected in the RAM direction. We choose five directions \citep{Zirnstein16} which largely avoid the region where the ribbon is present: Voyager 1 (centered on ecliptic J2000 coordinates of $\lambda,\beta=$-105$^{\circ}$, 35$^{\circ}$), port lobe (10$^{\circ}$, -12$^{\circ}$) defined as the left side of the heliosphere as it moves through the ISM as seen from within looking outward, downwind (75.7$^{\circ}$, -5.1$^{\circ}$; \citet{McComas15}), south pole (75.7$^{\circ}$, -90$^{\circ}$), and south lobe (89$^{\circ}$, -42$^{\circ}$). For the south lobe, the direction is based on the location of peak flux within the lobe. For the BU and Moscow models, we use different directions from that found in \citet{Zirnstein16} for IBEX observations. Due to the fact that the southern lobe only appears in the BU and Moscow models starting at the top three energy bands of IBEX-Hi, we average the location of peak flux within the southern lobe across the 1.74, 2.73, and 4.29 keV energy bands for each model. In the BU model, we take the south lobe direction to be ($\lambda,\beta$) =(69$^{\circ}$, -47$^{\circ}$). In the Moscow model, we take the south lobe direction to be ($\lambda,\beta$) =(67$^{\circ}$, -49$^{\circ}$). We calculate the ENA flux centered from regions centered on these directions spanning 15$^{\circ}$ in solid angle. The regions centered on these directions are presented in Figure \ref{fig:regions}.

In Figure \ref{fig:fluxcomp}, we present a comparison of the modeled ENA flux from the BU and Moscow models with IBEX observations averaged over the years 2009 through 
2013. Unlike in Figure \ref{fig:maps}, we do not scale the modeled ENA fluxes in Figure \ref{fig:fluxcomp}. For all five directions in Figure \ref{fig:fluxcomp}, the BU and Moscow fluxes are similar, though the ENA fluxes from IBEX observations are higher than modeled ENA fluxes. Most notably in the Voyager 1, port lobe, and downwind directions, the fluxes between the two models show good agreement, primarily at higher energies. For the south pole and south lobe directions, there is also good agreement between the models, though the BU and Moscow models show better agreement at low energies in the south lobe direction. 

A good method of interpreting IBEX data is by isolating the spectral slope ($\gamma$) of the ENA fluxes in a low energy bin and a high energy bin, such that $J(E) \propto E^{-\gamma}$, where $J(E)$ is the ENA flux and $E$ is the ENA energy \citep{Dayeh11,Dayeh14}. By separating the spectral slope from energies of 0.71 to 1.74 keV and 1.74 to 4.29 keV, we are able to focus on the contribution to the ENA flux from the slow solar wind plasma and the fast solar wind plasma, respectively. A comparison of spectral slopes for the different models and regions is presented in Table \ref{tab:gam}. We find good agreement in the port lobe direction between the BU and Moscow models at both the low and high energy bins. In the Voyager 1 direction we see a difference between the modeled spectral slopes of 18\% and 7\% at the low and high energy bins, respectively, caused by the slightly denser and hotter heliosheath plasma in the BU model. In the downwind direction we see a difference between the modeled spectral slopes of 7\% and 3\% at the low and high energy bins, respectively, demonstrating that the length of the low latitude heliotail does not significantly affect the modeled ENA results in the downwind direction. For the south pole direction, we note differences between the spectral slope for the BU and Moscow models of 32\% and 12\% for the low and high energy bins, respectively. In the south lobe direction, the differences for the low and high energy bins are 22\% and 9\%, respectively. In general, we find better agreement between the models at higher energies than at lower energies. In comparing with IBEX data, we generally do not find good agreement, which may be a result of the solar wind conditions used in the BU and Moscow models or the assumptions for the post-processed PUI distributions in the heliosheath. 

\begin{figure*}[t!]
\centering
  \includegraphics[scale=0.65]{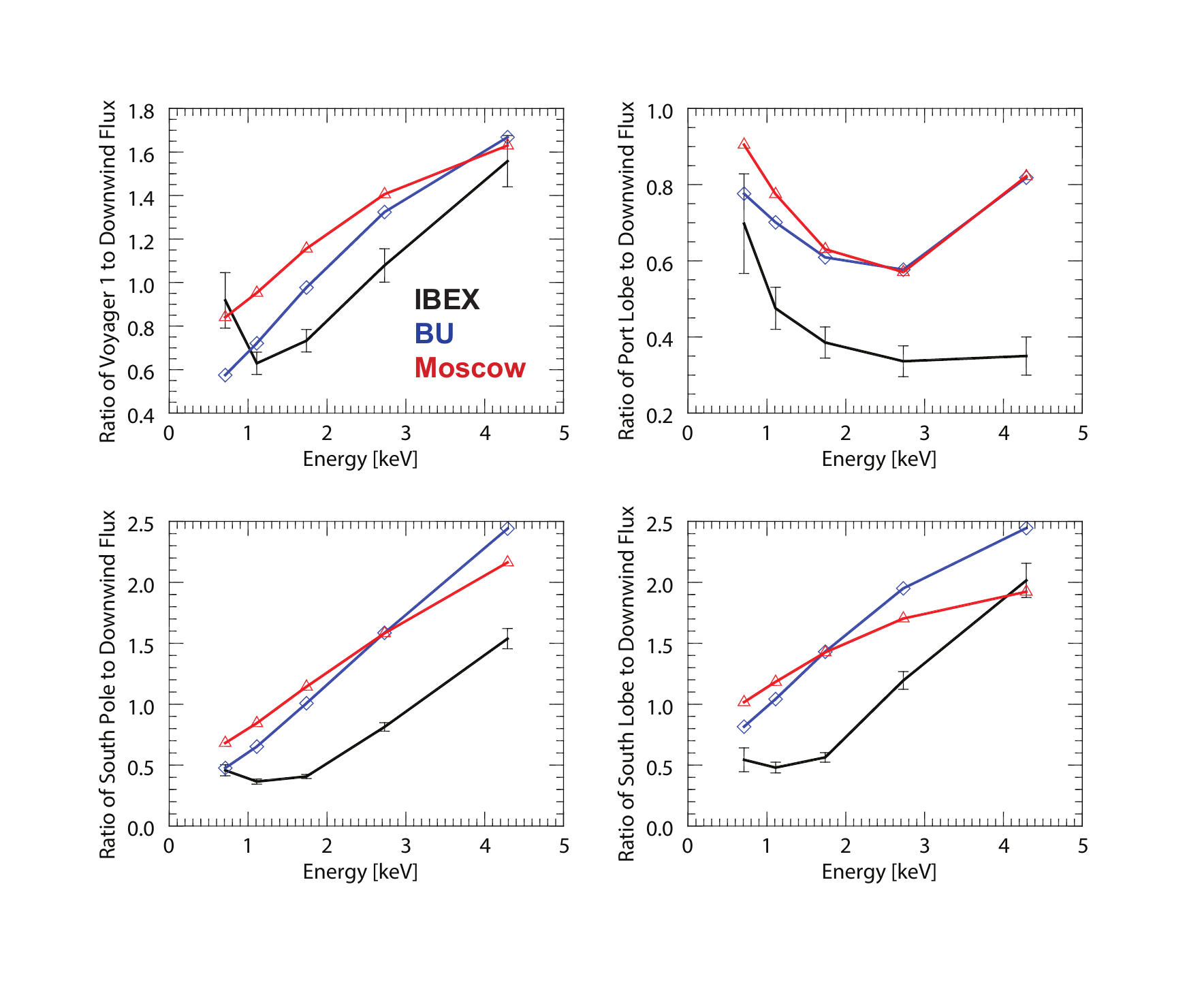}
  \caption{Ratios of ENA Flux for the Voyager 1 (top left), port lobe (top right), southern pole (bottom left), and southern lobe (bottom right) direction relative to ENA flux in the downwind direction. The black line corresponds to IBEX-Hi data averaged over the years 2009 through 2013, the blue line corresponds to the BU model, and the red line corresponds to the Moscow model. Fluxes are extracted over a 15$^{\circ}$ $\times$ 15$^{\circ}$ area centered around each direction.}
  \label{fig:ratios}
\end{figure*}

The overall agreement between the modeled fluxes demonstrates that the cooling length restricts the distance out to which IBEX observations are able to probe the length of the heliotail. Though there is overall agreement between the two models in terms of ENA flux, differences can still be noted in different regions of the heliosphere. Therefore, in order to directly compare with IBEX observations without the need to scale modeled results, we take ratios of the ENA flux for each direction relative to the downwind direction (Figure \ref{fig:ratios}). 

We choose the downwind direction as the reference point because unlike the Voyager 1 direction where the ENA flux is limited by the heliosheath thickness, which is overpredicted in MHD models and smaller than the cooling length, in the downwind direction the signal is only limited by the cooling length due to the thicker heliosheath. For the Voyager 1 and port lobe directions, the BU and Moscow models demonstrate similar trends to one another, though they deviate from one another at the lowest ENA energy and they do not agree with ratios from IBEX observations. For the south pole, the results between the two models are similar, with similar trends to the IBEX observations at energies above 1.74 keV, but with the ratio offset by a common factor of $\sim$1.5. For the south lobe, we note the largest differences. While the modeled ratios do not agree with the IBEX ratios, the BU and Moscow ratios also do not show good agreement between themselves. The south lobe has more flux at high energies in the BU model than in the Moscow model, which is reflected in this comparison. 

In Table \ref{tab:lin}, we present linear fits to the low-energy and high-energy portion of the ratios for each direction. Again, for the Voyager 1 direction we note that the BU and Moscow models show the best agreement at low energies. In the port lobe direction, the two models show best agreement at high energies. In the south pole and south lobe, differences are noted, where in both cases the BU model has a larger slope of the ratios at both the low and high energies than in the Moscow model. This demonstrates that at all energies the BU model has more ENA flux in the high latitude tail relative to the low latitude tail as compared to the Moscow model. In comparing the models with the IBEX data, the general trends agree, though quantitative agreement varies by region and energy range. While the BU and Moscow model do not show good agreement with the observations from the Voyager 1 direction at low energies, at high energies the results are in closer agreement. Conversely, in the port lobe the models show better agreement with the IBEX data at low energies than at high energies. For the south pole and south lobe, both models do not show good agreement with the data at low energies. At high energies for the south pole and south lobe, only good agreement is found between the Moscow model and data in the south pole; otherwise, the models do not compare favorably with the data.

One reason for the discrepancy between the modeled ratios and the IBEX ratios is the lack of inclusion of a time dependent solar wind in the MHD models. The MHD models using stationary solar wind conditions corresponding to an average over the 22-year solar cycle from 1995 to 2017, whereas the IBEX observations reflect an evolving heliosphere. Over the course of a solar cycle, the high latitude solar wind will vary between slow and fast solar wind. Based on the flow of the solar wind through the heliosphere, ENA observations along a given line of sight have the ability to probe both slow and fast wind. In the absence of solar cycle dependence in the models, the high latitudes are over-flooded with fast solar wind while minimizing the contribution from slow solar wind. Therefore, in a future study we will extend our investigation to include time dependent solar wind conditions. 

While the modeled ratios do show good agreement with respect to linear trends as shown in Figure \ref{fig:ratios} and Table \ref{tab:lin}, the percent differences between the modeled ratios must be greater than observational statistical limitations for the differences to be significant. In Table \ref{tab:ratio_diff}, we present the percent difference between the modeled ratios of BU and Moscow. While the ENA flux in the Voyager 1 direction for the two models is similar as has been noted prior, we find the greatest difference in the ratios relative to the downwind direction between the two models to be in the Voyager 1 direction. In this direction, we have the greatest difference between the BU and Moscow ratios of 46\% and 32\% at the 0.71 and 1.11 keV energy bands, respectively, falling below 20\% at the higher energy bands. For the south pole direction we find a difference of 25\% at the 0.71 keV energy band and in the south lobe direction we find a 21\% difference at 4.29 keV, but otherwise the differences for the remaining directions and energy bands fall below 20\%. In the south lobe direction, we find differences of 43\% and 29\% for the 0.71 and 1.11 keV energy bands, and the higher energy bands having differences less than 20\%. This is important as the absolute uncertainty of the IBEX-Hi energy spectra is estimated to be $\sim$20\% \citep{Fuselier12}. Therefore, in general the differences between the ratios of the two models above 2 keV are not statistically significant enough for IBEX observations to distinguish different heliotail configurations.

\section{Conclusions}\label{sec:Summary}
\begin{figure*}[t!]
\centering
  \includegraphics[scale=0.5]{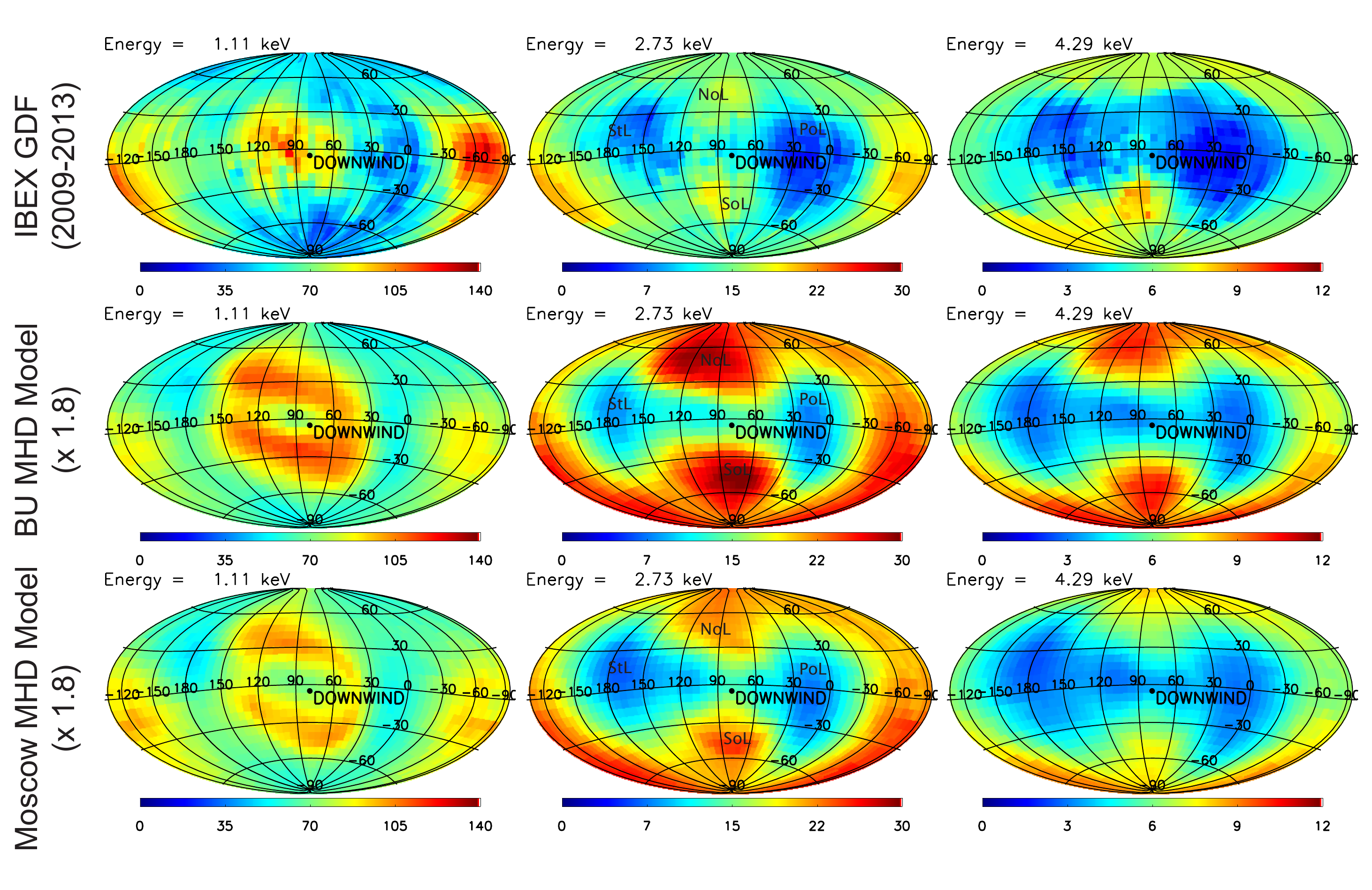}
  \caption{ENA sky map of flux centered on the downwind (tail) direction in units 
  of (cm$^{2}$ s sr keV)$^{-1}$ of the IBEX GDF from the first five years (2009-2013) of observations (top row), BU model (middle row), and 
  Moscow model (bottom row). The energies included are 1.11 (left column), 2.73 (middle column), and 4.29 keV (right column). Simulated sky maps are 
  multiplied by a factor of 1.8. Labels on maps at 2.73 keV are for the north lobe (NoL), south lobe (SoL), port lobe (PoL), and starboard lobe (StL).}
  \label{fig:maps}
\end{figure*}

\begin{figure*}[t!]
\centering
  \includegraphics[scale=0.55]{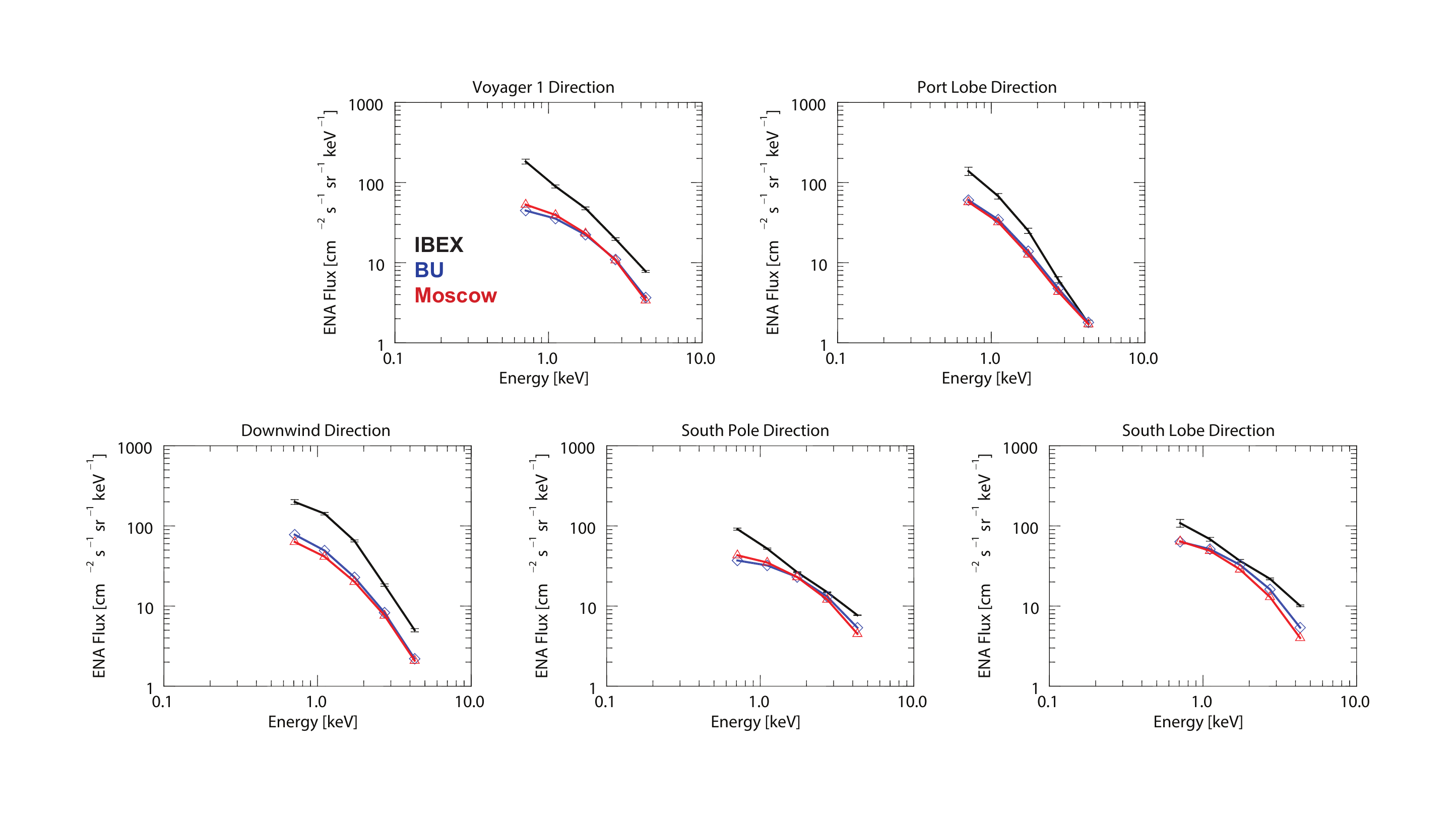}
  \caption{ENA flux spectra for the directions of Voyager 1 (top left), the port lobe (top right), downwind (bottom left), the southern pole (bottom middle), and the southern lobe (bottom right), in units of [cm$^{-2}$ s$^{-1}$ sr$^{-1}$ keV$^{-1}$]. The black line corresponds to IBEX-Hi data averaged over the years 2009 through 2013, the blue line corresponds to the BU model, and the red line corresponds to the Moscow model. Simulated ENA fluxes are not scaled and fluxes for all cases are extracted over a 15$^{\circ}$ $\times$ 15$^{\circ}$ area centered around each direction.}
  \label{fig:fluxcomp}
\end{figure*}

The ENA model fluxes produced by using the same prescription, but based on two different plasma and H atom solutions obtained from the BU and Moscow models, are generally in good agreement, though some differences are still observed. The cooling length places a limit on how far ENAs can be observed at a particular energy band, and at the IBEX-Hi energies the configuration of the heliotail is unable to be distinguished between the two models at most energies. Both solutions when using the same ENA prescription are unable to quantitatively reproduce IBEX ENA observations without scaling the results up by a factor approximately 1.8, although this does not necessarily apply to all directions in the sky (it is clear in Figures \ref{fig:maps} and \ref{fig:fluxcomp} the scaling sometimes needs to be higher, sometimes lower). There are slight differences between the modeled ENA fluxes, which arise from the role of non-ideal MHD processes compressing the heliosphere at high latitudes in the BU model with respect to the Moscow model \citep{Kornbleuth21}. 

We investigate differences and similarities between the two models. We note similar ENA spectra in the Voyager 1 and port flank directions for the models and small differences between the spectra in the tail, south pole, and south lobe directions. We also compare ratios of the ENA flux for each given direction relative to the ENA flux in the tail direction and note that the general trend between the modeled and observed ENA ratios agree. While we find differences in these ratios primarily in the south pole and south lobe directions, we find that the difference in these ratios may not be sufficiently large to be observed by IBEX considering the estimated $\sim$20\% systematic uncertainty of IBEX-Hi observations \citep{Fuselier12}. 

While differences resulting from the different heliotail configurations may not be discernible at the IBEX-Hi energies, at higher energies it may be possible to distinguish between the two different heliotails due to the smaller charge exchange cross section \citep{Lindsay05} and thus longer cooling lengths. In future studies, we will investigate and compare simulated ENA maps from the BU and Moscow models at the INCA energies (5.2 to 55 keV) \citep{Krimigis09,Dialynas17} and at the IMAP-Ultra energies (up to 300 keV) \citep{McComas18} in order to determine at which energies the shape of the heliotail can be discerned.

\acknowledgments
The authors were supported by NASA grant 18-DRIVE18$\_$2-0029, Our
Heliospheric Shield, 80NSSC20K0603. The work of I.B. and V.I. was conducted in the framework of a topic of the state assignment ``Plasma'' to the Space Research Institute, Russian Academy of Sciences. The work at JHU/APL was supported also by NASA under contracts NAS5 97271, NNX07AJ69G, and NNN06AA01C and by subcontract at the Office for Space Research and Technology. Resources supporting this work were provided by the NASA High-End Computing (HEC) program through the NASA Advanced Supercomputing (NAS) Division at Ames Research Center. The authors would like to thank the staff at NASA Ames Research Center for the use of the Pleiades super-computer under the award SMD-20-46872133.


\newpage

\begin{table}[t!]
\centering
\begin{tabular}{cccc}
\hline
    Region          & Case          & $\gamma_{1}$     & $\gamma_{2}$     \\
                    &               & (0.71 - 1.74 keV)& (1.74 - 4.29 keV)\\
\hline
 Voyager 1          & BU            & 0.78             & 2.00             \\
                    & Moscow        & 0.92             & 2.13             \\
                    & IBEX          & 1.48 $\pm$ 0.09  & 2.01 $\pm$ 0.06  \\
 Port Lobe          & BU            & 1.64             & 2.27             \\
                    & Moscow        & 1.68             & 2.22             \\
                    & IBEX          & 1.97 $\pm$ 0.15  & 2.96 $\pm$ 0.14  \\
 Downwind           & BU            & 1.37             & 2.59             \\
                    & Moscow        & 1.28             & 2.51             \\
                    & IBEX          & 1.42 $\pm$ 0.07  & 2.84 $\pm$ 0.06  \\
 South Pole         & BU            & 0.53             & 1.61             \\
                    & Moscow        & 0.70             & 1.81             \\
                    & IBEX          & 1.42 $\pm$ 0.03  & 1.39 $\pm$ 0.02  \\
 South Lobe         & BU            & 0.74             & 2.00             \\
                    & Moscow        & 0.90             & 2.18             \\
                    & IBEX          & 1.28 $\pm$ 0.11  & 1.49 $\pm$ 0.05  \\
\hline
\end{tabular}
\caption{Spectral slopes from the BU and Moscow models. as well as IBEX data averaged over the years 2009 - 2013 in the directions of Voyager 1, the port lobe, downwind, the south pole, and the south lobe. Spectral slopes are calculated using fluxes averaged over a
15$^{\circ}$ $\times$ 15$^{\circ}$ area centered around each direction. $\gamma_{1}$ is the spectral slope over the low-energy portion of the spectrum, 0.71-1.74 keV. 
$\gamma_{2}$ is the spectral slope over the high-energy portion of the spectrum, 1.74-4.29 keV.}
\label{tab:gam}
\end{table}

\begin{table}[t!]
\centering
\begin{tabular}{cccc}
\hline
    Region          & Case          & $m_{1}$     & $m_{2}$     \\
                    &               & (0.71 - 1.74 keV)& (1.74 - 4.29 keV)\\
\hline
 Voyager 1          & BU            & 0.39             & 0.27             \\
                    & Moscow        & 0.31             & 0.18             \\
                    & IBEX          & 0.03 $\pm$ 0.10  & 0.33 $\pm$ 0.05  \\
 Port Lobe          & BU            & -0.16            & 0.09             \\
                    & Moscow        & -0.26            & 0.08             \\
                    & IBEX          & -0.20 $\pm$ 0.09 & -0.01 $\pm$ 0.03 \\
 South Pole         & BU            & 0.52             & 0.56             \\
                    & Moscow        & 0.45             & 0.40             \\
                    & IBEX          & 0.01 $\pm$ 0.04  & 0.43 $\pm$ 0.03  \\
 South Lobe         & BU            & 0.60             & 0.39             \\
                    & Moscow        & 0.40             & 0.19             \\
                    & IBEX          & 0.08 $\pm$ 0.08  & 0.59 $\pm$ 0.05  \\
\hline
\end{tabular}
\caption{Linear fits to the ratios of ENA flux for different directions relative to the downwind direction for the BU and Moscow models. as well as IBEX data averaged over the years 2009 - 2013 in the directions of Voyager 1, the port lobe, the south pole, and the south lobe. Linear fits are calculated using ratios from fluxes averaged over a 15$^{\circ}$ $\times$ 15$^{\circ}$ area centered around each direction. $m_{1}$ is the linear slope over the low-energy portion of the spectrum, 0.71-1.74 keV. $m_{2}$ is the linear slope over the high-energy portion of the spectrum, 1.74-4.29 keV.}
\label{tab:lin}
\end{table}

\begin{table}[t!]
\centering
\begin{tabular}{cccccc}
\hline
    Region          & 0.71 keV  & 1.11 keV  & 1.74 keV  & 2.73 keV  & 4.29 keV \\
\hline
 Voyager 1          & 46.2\%    & 32.2\%    & 18.3\%    &  6.2\%    &  2.3\%   \\
 Port Lobe          & 16.6\%    & 10.6\%    &  3.5\%    &  1.1\%    &  0.5\%   \\
 South Pole         & 24.7\%    & 13.5\%    &  0.4\%    &  12.7\%    & 21.4\%   \\
 South Lobe         & 43.1\%    & 29.4\%    &  13.5\%\   &  0.2\%    & 11.4\%   \\
\hline
\end{tabular}
\caption{Percent Difference between BU and Moscow model ratios of ENA flux in different regions with respect to the downwind direction ($|$BU-Moscow$|$/BU). Included are the directions of Voyager 1, the port lobe, the south pole, and the south lobe for IBEX-Hi energies ranging from 0.71 keV to 4.29 keV.}
\label{tab:ratio_diff}
\end{table}

\end{document}